# Design and Performance Analysis of a 2.5 MW-Class HTS Synchronous Motor for Ship Propulsion

Jin Zou, Di Hu, Tim J. Flack, Haichao Feng, Xiaozhuo Xu, and Mark D. Ainslie, *Member, IEEE*

*Abstract*--The development of cryogenic technology and high temperature superconducting (HTS) materials has seen continued interest worldwide in the development of HTS machines since the late 1980s. In this paper, the authors present a conceptual design of a 2.5 MW class synchronous motor. The structure of the motor is specified and the motor performance is analyzed via a three-dimensional model using the finite element method (FEM). Rotor optimization is carried out to decrease the harmonic components in the air gap field generated by HTS tapes. Based on the results of this 3D simulation, the determination of the operating conditions and load angle is discussed with consideration to the HTS material properties. The economic viability of air-core and iron-core designs is compared. The results show that this type of HTS machine has the potential to achieve an economic, efficient and effective machine design, which operates at a low load angle, and this design process provides a practical way to simulate and analyze the performance of such machines.

## I. Introduction

WITH the technological advancement of High Temperature Superconductors (HTS), advanced synchronous rotating machines, like 'SuperMachines' [1], have been designed, constructed and tested. These machines utilize the advantageous properties of superconducting materials, such as zero resistance and higher current density [2]. Hence, a superconducting machine has advantages over a conventional design, such as lower power losses, more compact volume, and higher output power and output torque [3].

Conventional machines for ship propulsion to provide high electric power have been developed and optimized for decades, but these machines cannot achieve multi-megawatt levels without paying a significant penalty in size and efficiency [4]. An HTS superconducting machine can have a reduction as much as half the weight of a conventional machine for the same rated power [5]. In addition, when the rating is over 1000 hp, the HTS machine offers the opportunity to reduce losses by half [6]. It is typical that the power generation subsystem for seaborne platforms must be packaged with limited space and weight [4]. For military ships, the lower weight of the ship's body ensures higher speed and more space for cargo and equipment, which in turn guarantees enhanced capabilities. Employed with an HTS superconducting motor, a commercial ship has more space to carry more passengers and goods, and therefore its economic value is improved. Thus, the HTS superconducting motor is the ideal machine for next-generation ship propulsion applications.

In this paper, a 2.5 MW class synchronous motor is designed for ship propulsion, which requires for a low rotational speed and high thrust force. The motor performance is analysed via a three-dimensional (3D) model using the finite element method (FEM). The paper is organised as follows: in Section 2, the structure of the HTS motor is described, including the stator and rotor topologies, as well as the motor specification and rationale behind the design choices. Optimization of the rotor geometry is also carried out to generate a more sinusoidal waveform in the air gap. In Section 3, considering the technical challenges in HTS machine design, the operating conditions (including operating current and temperature) of the motor are determined, the amount of HTS conductors saved by inserting an iron core into the design is highlighted and an analysis of the motor performance is also presented. Finally, the advantages of this HTS machine design, compared with conventional copper machine are emphasized in Section 4.

D. Hu and J. Zou would like to acknowledge financial support from Churchill College, the China Scholarship Council and the Cambridge Commonwealth, European and International Trust. M. D. Ainslie would like to acknowledge financial support from a Royal Academy of Engineering Research Fellowship. This work is supported by a Henan International Cooperation Grant: 144300510014.

D. Hu, J. Zou and M. D. Ainslie are with the Bulk Superconductivity Group, Department of Engineering, University of Cambridge, Cambridge CB2 1PZ, UK (e-mail: dh455@cam.ac.uk, jz351@cam.ac.uk, mda36@cam.ac.uk).

T. J. Flack is with the Electrical Engineering Division, Department of Engineering, University of Cambridge, CAPE Building, Cambridge CB3 0FY, UK (e-mail: tjf1000@cam.ac.uk).

X. Xu and H. Feng are with the School of Electrical Engineering and Automation, Henan Polytechnic University, Jiaozuo, Henan 454000, P.R. China (e-mail: xxz@hpu.edu.cn, fhc@hpu.edu.cn)

## II. HTS MOTOR STRUCTURE SPECIFICATION

The HTS motor configuration is shown in Figure 1, which consists of a stator yoke, armature winding, rotor iron core, HTS field winding and field winding support structure. Conventional copper winding is utilised in the stator with some modification to maximize the air gap magnetic field. HTS tape in the form of a racetrack coil is used as the field winding. The machine design specification is shown in Table I.

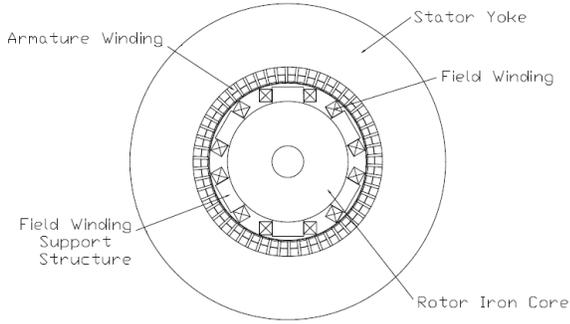

Fig.1. HTS machine design with copper armature winding and HTS field winding

TABLE I Machine design specification

| Pole pairs number | 3 |
|---|---|
| Rated speed | 480 rpm |
| Stator turn number per phase | 630 |
| Stator slot number | 54 |
| Rotor turn number per pole | 1200 |
| Effective length | 600 mm |
| Stator outer diameter | 1000 mm |
| Stator inner diameter | 500 mm |
| Air gap | 3.5 mm |
| Average air gap field | ~1 T |

### A. Rotor Topology

Although there are advantages to an air-core structure, such as lower current, lower moment of inertia, and less noise [7], a more robust winding support structure must be designed to transmit the torque of the motor. Furthermore, an iron core can help generate more flux in the air gap, and the inserted magnetic core in the rotor can effectively reduce the Ampere turns required for the air gap flux density. Considering the high cost of HTS tape [8], the reducing amount of ampere turns can decrease the cost of an HTS machine significantly, which improves the economic competitiveness with the conventional motors. This will be discussed in detail in Section 3. Although compared with air-core design the weight advantage of an HTS machine is limited, in this design its high efficiency and high overload capability still justifies its superiority over conventional induction/synchronous machines in Section 3.

### B. Stator Topology

The stator design is similar to a conventional motor, whereby a magnetic stator yoke is employed in the motor to guide the magnetic flux, but with nonmagnetic teeth and higher Ampere turn loading (see Fig. 2) [7]. Short-pitch, two-layer winding is utilized in this HTS motor design, reducing the voltages of the phase windings and the amount of copper used. Another advantage of employing a short-pitched winding is that it can generate a more sinusoidal current linkage distribution than for a full-pitched winding. A copper stator allows us to avoid complications with AC loss considerations of HTS materials that can dramatically affect motor performance [9]-[11], such as decreasing motor efficiency and increasing cooling system cost and complexity.

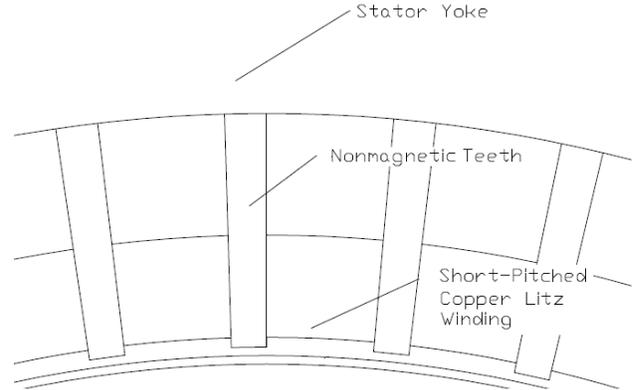

Fig.2. Toothless stator design with copper Litz winding

Considering the limitation on the flux density due to saturation of the teeth, the conventional iron teeth configuration is replaced by nonmagnetic teeth [7]. As a consequence, the efficiency will be decreased somewhat due to an increase in leakage flux, but a higher flux density and higher voltage can be achieved. In addition, compared with its conventional counterpart (93%) [5], this type of 2.5 MW HTS motor has higher efficiency (over 97%), as calculated in Section 3. These features are more attractive for high torque machines for maritime propulsion. G10 FRP (Fibre Reinforced Plastic), which has the same permeability as air [12], is utilized in the simulation model to support the armature winding. Nonmagnetic teeth with smaller width can benefit the omission of slot harmonics, but increase the transverse flux components [7]. Therefore, smaller diameter copper strands (Litz wire) are utilised in this design to decrease eddy current losses. In addition, there is a certain teeth width required in magnetic teeth design to stay safely with tolerable saturation level. With nonmagnetic teeth a smaller width of the teeth can be applied, thus a higher Ampere-turn loading is achieved in this design.

### C. FEM modeling specification

Infolytica's MagNet is software that uses FEM techniques to simulate static, frequency dependent or time-varying electromagnetic fields, and is especially suitable for electric machine simulation in both 2D and 3D. The automated features of MagNet reduce the time required for the design cycle significantly [13]. This 2.5 MW HTS machine design is based on the geometry shown in Fig. 1. Copper is used in the armature windings in the model with an electrical conductivity of $5.9 \times 10^7$ S/m. To simulate the zero resistance property of the HTS coils, the copper with an electrical conductivity of $5.9 \times 10^{80}$ S/m is used to represent superconducting properties in the field winding. The geometry of the 3D model is presented in Fig. 3 and the parameters used for the materials are shown in Table II.

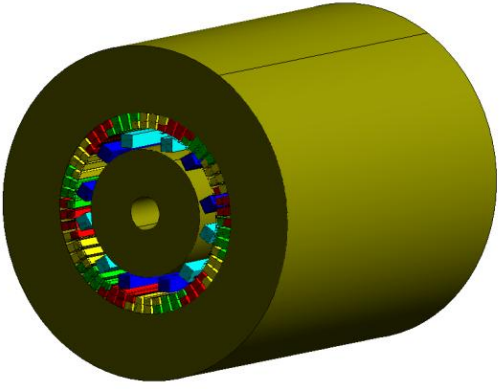

Fig.3. Geometry of the 3D model built in MagNet

TABLE II Parameters used in FEM model

| Material | Magnetic permeability | Electrical conductivity | Electrical permittivity |
|---|---|---|---|
| HTS | $\mu_0$ ($1.26 \times 10^{-6}$) | $5.9 \times 10^{80}$ S/m | $\varepsilon_0$ ($8.85 \times 10^{-12}$ F/m) |
| Iron (without losses) | Shown in Fig. 4 | 0 | $\varepsilon_0$ ($8.85 \times 10^{-12}$ F/m) |
| Copper | $\mu_0$ ($1.26 \times 10^{-6}$) | $5.9 \times 10^7$ S/m | $\varepsilon_0$ ($8.85 \times 10^{-12}$ F/m) |
| Air | $\mu_0$ ($1.26 \times 10^{-6}$) | 0 | $\varepsilon_0$ ($8.85 \times 10^{-12}$ F/m) |

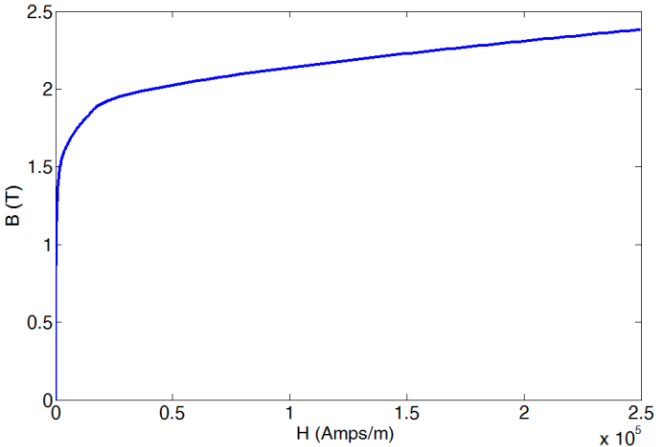

Fig.4. Magnetic permeability of the iron material used in the 3D model

### D. Rotor optimization

Compared with the harmonics generated in the air gap of a conventional copper machine, which normally has a 5th harmonic of about 8% [1], the harmonics in the air gap field in the HTS machine designed in this paper are relative low, as shown in Fig.5.

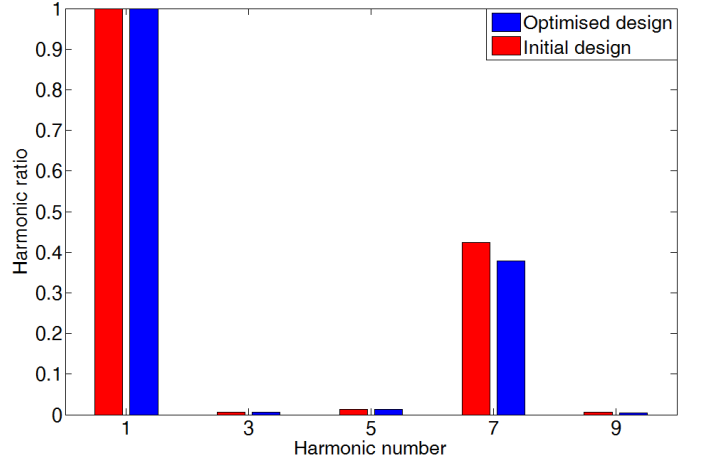

Fig.5. Harmonic ratios of odd harmonics of the air gap field

Although the harmonics are already low in the air gap, especially the 3rd and 5th harmonics, it cannot be ignored that the 7th harmonic occupies 27% of the whole waveform, which causes the fluctuation of torque, especially in high rotation [14]. Therefore based on our previous rotor optimization method [14], the rotor geometry is optimized using the 3D model to further decrease the higher-order harmonics in the air gap. Based on this optimization, in one pole pair, the width of coil becomes 65.2 mm, the distance between the coil sides is 283 mm.

In Fig. 5, harmonic ratio of each harmonic in optimized design is lower than that in the initial 3D model. Especially for 7th harmonics, the improvement is obvious. Total harmonic distortion (THD) of the air gap field to evaluate the effectiveness of harmonics optimization is calculated following equation (1)

$$\text{THD} = \frac{\sqrt{\sum_{n=2}^{\infty} B_{rn}^2}}{B_{r1}} \quad (1)$$

In (1), $B_{rn}$ represents the $n^{th}$ harmonic of the perpendicular magnetic field. With the parametric optimization, THD decreases from 0.42 to 0.38 in our optimized design, which again verifies that the optimization was successful.

### III. OPERATION SPECIFICATION

## A. Operating point determination

2G-HTS (YBCO) tape is chosen as the conductor in the field winding, whose parameters are depending on their peripheral temperature and magnetic field. Therefore, it is necessary to choose a suitable operating temperature to satisfy the design requirements. Since the manufacturer provides the dependence of critical current density on magnetic fields for various temperatures of HTS tapes based on the one at 77 K, the analysis of the operating point starts from the specification at 77 K, which is shown in Table III to the performance at different temperatures (see Fig. 6), as per currently available commercial tape options.

TABLE III HTS specification

| Minimum critical current | Critical current density | Width | Total wire thickness |
|---|---|---|---|
| 80 A | $2 \times 10^8$ A/cm$^2$ [self-field, 77 K] | 4 mm | 0.1 mm |

At 77 K in self-field (i.e., transport current only with no external magnetic field), the critical current is 80 A, but under different background fields and different angles between the field and the HTS tape surface, the critical current will decrease. In the tape under consideration in this analysis, it is assumed as an approximation that the critical current is reduced most significantly due to a field perpendicular to the tape surface. This allows employment of a Kim-like model [15], although this is not necessarily true for all 2G YBCO tapes [16]-[18]. The in-field critical current characteristics are shown in Fig. 6 [18].

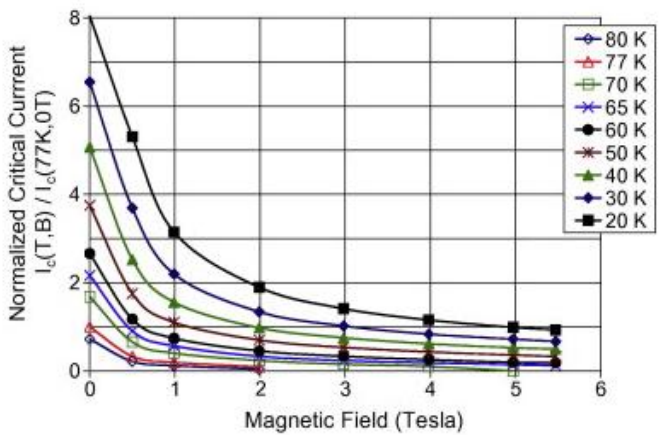

Fig. 6. The dependence of the critical current on magnetic field for various temperatures with the field oriented perpendicular to the ab-plane of the HTS tape [18]

Considering the required 1 T average air gap field in the design specification of Table III and based on the results in this 3D model, the maximum perpendicular flux density around the HTS coils is 1.6 T. This flux density profile of the motor is shown in Fig. 7. The geometry of Fig. 7 is exactly the same as that of Fig. 1.

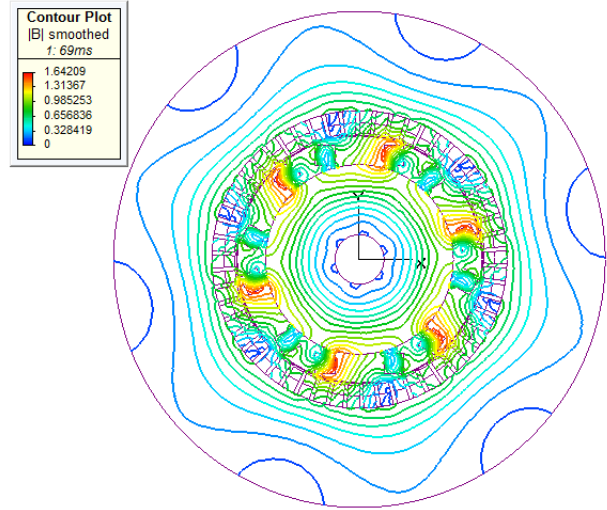

Fig. 7. Overall HTS motor flux density profile

According to Fig. 6, if the temperature is as high as 77 K, the critical current will be too low to be practical under this magnitude of perpendicular field, so that a much larger quantity of HTS tape is required. However, the cooling system would be cheaper and less complicated than that of lower temperatures.

If the motor were to be operated at 40 K, according to Figure 6 with a perpendicular magnetic field 1.6 T, the critical current density decreases, but is still approximately 1.2 times of that at 77 K in self-field. Based on Fig. 6 and Table III, at 40 K the critical current at 1.6 T would be 96 A. In order to avoid quenching of the coils in the superconducting field winding, a proper safety margin for the current is chosen: 80% of the critical current [19]. Based on an operating temperature of 40 K, the designed motor will run with an excitation current of 75 A [8]. The specification for the motor's operation point can be found in Table IV.

TABLE IV HTS synchronous motor specification

| Maximum field coil flux density | 1.6 T |
|---|---|
| HTS tape requirement | ~ 8.6 km |
| Field coil excitation current | 75 A |
| Stator terminal voltage | 6000 V |
| Nominal output power | 2.5 MW |
| Open-circuit terminal voltage | 5805 |

## B. Load angle determination

The determination of the load angle depends on the maximum torque, which is obtained at 90 electrical degrees. If the load angle is greater than 90 degrees, the motor will lose synchronism. The load angle characteristics of this 2.5 MW HTS motor are shown in Figs. 8-10.

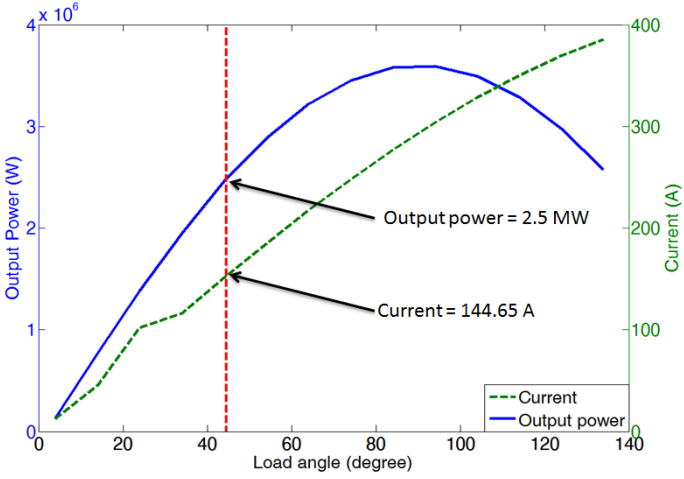

Fig. 8. Output power and current for different load angles

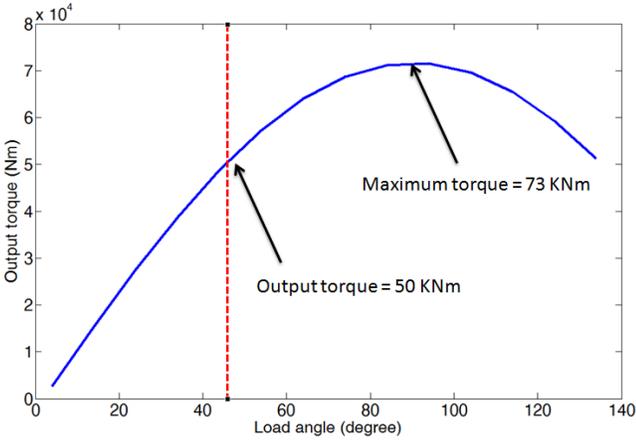

Fig. 9. Nominal torque and maximum torque for different load angles

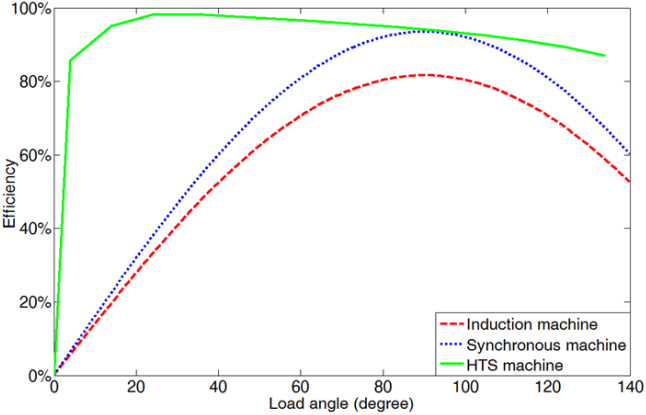

Fig.10. Efficiency of HTS motor for different load angles compared with conventional machines

When the load angle is 90 degrees, the output torque is 72 kNm and the output power is 3.6 MW. Considering the load angle margin, the nominal torque is set as two-thirds of the maximum torque. The output torque and power are 50 kNm and 2.5 MW, respectively, with a corresponding load angle of 44 degrees. When the HTS motor operates at a load angle of 44 degrees, the efficiency (excluding the cooling system loss) is around 97.6%, with a power factor of 0.9.

In Fig. 10, compared with traditional induction and synchronous machines, the HTS machine can maintain high efficiency over a much larger range of load angles. Furthermore, the overall efficiency is much higher when compared with the traditional machines.

When the load angle is set as 44 degrees, the magnetic field profile of this HTS machine is shown in Table V.

TABLE V Flux density around the HTS field coil

| $B_{perp}$ | $B_{para}$ | $|\mathbf{B}|$ |
|---|---|---|
| 1.2T | 0.78T | 1.4 T |

In Table V, $B_{perp}$ and $B_{para}$ are the magnitudes of the perpendicular and parallel fields, respectively, where

$$|\mathbf{B}| = \sqrt{B_{perp}^2 + B_{para}^2} \quad (2)$$

Based on Table V, the final design has a perpendicular magnetic field density of 1.2 T and a total flux density of 1.4 T around the HTS coils, which are both smaller than 1.6 T. In addition, considering the saturation magnetic field of iron as around 1.7 T, these results satisfy the requirement that maximum magnetic field is no more than 1.6 T. The HTS machine therefore operates within safe operating limits.

### C. Iron core insertion design

Since, at current prices, the HTS conductor cost usually comprises a large portion of total machine cost [8], decreasing the amount of HTS material employed in the machine is significant. Therefore, it is critical to seek methods to decrease the amount of HTS material used, whilst maintaining the same rated output and high efficiency in the machine design.

Compared with the air-core design, the iron core insertion design used in this paper can decrease the Ampere turns, and therefore decrease the length of HTS material used in the machine. For the same output power and average air gap field with the iron-core insertion design, which employs 1200 turns of HTS tape (8.6 km in length), the air-core design requires for approximately 1800 turns of HTS conductor (13 km in length, approximately 1.5 times the length used in the design with the iron-core insertion). Therefore this HTS machine design has the potential to optimize the cost to two thirds of its original capital investment, which will save hundreds of thousands of pounds at current HTS tape prices.

Although a further reduction in size of HTS machines is achievable by air-core design, iron core is considered by machine designers if the focus is on optimizing the amount of required HTS wire, with reasonable reduction in the volume.

## IV. ADVANTAGE OF THIS MACHINE DESIGN

Compared with conventional electric machines of this type, the HTS machine designed in this paper achieves optimization in the following aspects:

Firstly, it improves the stator design by utilizing a special stator configuration with nonmagnetic teeth. Compared with conventional copper stators, in which the tooth width is limited to a certain level to remain within a tolerable saturation level, the stator with nonmagnetic teeth (see Fig. 2) can further decrease the tooth width to increase the conductor cross-section in the armature winding. Thus, a higher Ampere-turn loading can be achieved with more conductors in the stator slots.

In addition, although the harmonic components in the air gap field of the HTS machine are relatively low, they are further reduced in this paper by geometrical optimization of the rotor design.

High machine efficiency is achieved, and since this 2.5 MW class HTS machine can operate at a very small load angle, the overload capability is improved significantly. Due to the small load angle, switching on/off has little influence on the voltage. Furthermore with the small load angle, the tilting moment of the HTS machine is enlarged compared with conventional machines [20].

The economics of the machine are also considered. With the iron-core insertion design, one third of the cost of the HTS tape required is saved, which increases the economic viability of this machine design.

Therefore, this HTS machine design provides optimization in the stator utilization level, harmonic components in air gap field, machine efficiency, overload capability and economic liability.

## V. CONCLUSION

Progress in HTS materials and cooling systems has helped open the door for the development of HTS machines for industrial applications. A conceptual design of a 2.5 MW synchronous HTS motor that achieves a torque of 50 kNm at 480 rpm is described in this paper. The specific structure of this 2.5 MW HTS motor is also discussed. An iron core is inserted back into the rotor, in order to generate higher flux with relatively lower Ampere-turns. This configuration helps to decrease the amount of HTS tape used in the design, which increases its economic viability when compared with a conventional motor. A method to determine the operating point and load angle, in consideration of the HTS material properties, is also provided. The design procedure described in this paper provides a practical way to simulate and analyse the performance of such HTS machines.

## VI. ACKNOWLEDGEMENT

The authors would like to thank Prof. Archie Campbell, and Drs. Patrick Palmer and Tim Coombs for their helpful discussion on the original version of this document.

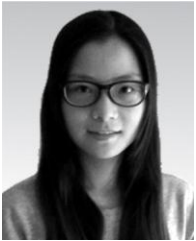
**Jin Zou** was in Henan Province, China, in 1989. She received the B.Eng. from both University of Birmingham, UK and Central South University, China from in 2012. Now she is the PhD student in University of Cambridge. Her main research interests include property analysis of High Temperature Superconducting bulk material and the simulation and control of High Temperature Superconducting machine.

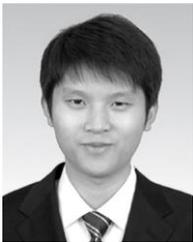
**Di Hu** was born in Wuhan, China, in 1990. He received the B. Eng degree in electronic and electric engineering from both Huazhong University of Science and Technology and University of Birmingham in 2012, and is currently working toward the Ph.D. degree in engineering at University of Cambridge, Cambridge, UK

His areas of interests include the theoretical analysis, 2-D, 3-D finite element method simulation of High Temperature Superconductor (HTS) machine and properties research of HTS tape, bulk for electric machines.

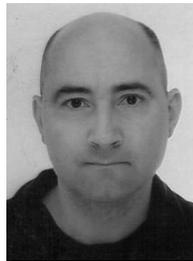
**Tim Flack** was awarded his B.Sc in Electrical Engineering from Imperial College, London, in 1986 and obtained his PhD from the same institution in 1990. Since then he has worked as a post-doc, and then as a University Lecturer at the Cambridge University Engineering Department.

His research interests include: the development of 2-D and 3-D finite-element codes for the modelling of electrical machines and also nano and micromagnetic simulations; electromagnetically-geared motors and generators; applied superconductivity and its modelling using the finite-element method; development of the brushless doubly-fed generator for wind power.

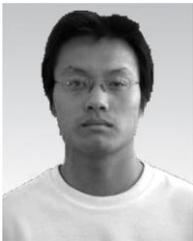
**Haichao Feng** was born in China in 1983, and received the B.S. and M.S. degrees in electrical engineering and automation, control theory and control engineering from the School of Electrical Engineering and Automation, Henan Polytechnic University, China, in 2005 and 2008, respectively. He is currently a teacher in the School of Electrical Engineering and Automation, Henan Polytechnic University, China.

Mr. Feng is a member of the Institute of Linear Electric Machines and Drives, Henan Province, China.

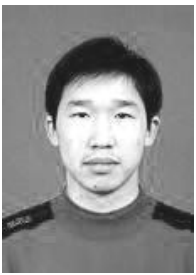
**Xiaozhuo Xu** was born in China in 1980, and received the B.S. and M.S. degrees in electrical engineering and automation, motor and electrical from the School of Electrical Engineering and Automation, Henan Polytechnic University, China, in 2003 and 2006, respectively. His research interests are the analysis of physical fields for special motors, and the optimization design of linear and rotary machines.

He is currently a lecturer in the School of Electrical Engineering and Automation, Henan Polytechnic University, China.

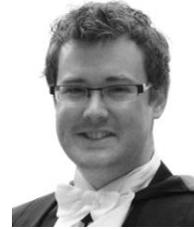
**Mark Douglas Ainslie** (M'2005) was born in Adelaide, Australia, in 1981. He received the B.E. & B.A. (Japanese) degree from the University of Adelaide, Adelaide, Australia, in 2004, the M.Eng. degree from the University of Tokyo, Tokyo, Japan, in 2008, and the Ph.D. degree from the University of Cambridge, Cambridge, UK, in 2012.

He is currently a Royal Academy of Engineering Research Fellow in the Bulk Superconductivity Group at the University of Cambridge and is a Junior Research Fellow at King's College, Cambridge. His current research interests are in applied superconductivity in electrical engineering, including superconducting electric machine design, power system protection and energy storage, electromagnetic modeling, and interactions between conventional and superconducting materials.